\documentclass[aps,showpacs,preprintnumbers]{revtex4}
\usepackage{amsmath}
\usepackage{graphicx}
\renewcommand{\Re}{{\rm Re}\;}
\newcommand{\Sp}{{\rm Sp}\;}
\newcommand{\fNR}{f_{\rm NR}}

\newcommand{\Was}{W\c as}
\newcommand{\KKMC}{\hbox{${\cal KK}$}\ MC}
\newcommand{\Eq}[1]{eq.~(\ref{#1})}
\newcommand{\Ref}[1]{ref.~\cite{#1}}
\newcommand{\Refs}[1]{refs.~\cite{#1}}
\newcommand{\Li}{{\rm Li}}
\newcommand{\Lf}{{\rm Lf}}
\newcommand{\Sf}{{\rm Sf}}

\newcommand{\MISR}[2]{{\cal M}^{\rm ISR (#2)}_{#1}}
\newcommand{\sISR}[2]{{\sigma}^{\rm ISR (#2)}_{#1}}
\newcommand{\BYFS}{B_{\rm YFS}}
\begin{document}
\title{Comparisons of Exact Results for the Virtual Photon Contribution to 
Single Hard Bremsstrahlung in Radiative Return for 
$e^+e^-$ Annihilation}\thanks{Work partly supported by Department of Energy 
grant DE-FG02-05ER41399 and NATO grant PST.CLG.980342.}
\author{S. Jadach}
\affiliation{CERN, Theory Division, CH-1211 Geneva 23, Switzerland,\\
Institute of Nuclear Physics, ul. Radzikowskiego 152, Krak\'ow, Poland}
\author{B.F.L. Ward}
\affiliation{Department of Physics, Baylor University\\
  One Bear Place \#97316, Waco, Texas 76798-7316, USA}
\author{S.A. Yost}
\affiliation{Department of Physics, Baylor University\\
  One Bear Place \#97316, Waco, Texas 76798-7316, USA}
\pacs{12.20.Ds\hfill Submitted to Phys.\ Rev.\ D}
\preprint{BU-HEPP 05/03}
\date{February 21, 2006}
\revised{March 20, 2006}
\begin{abstract}
We compare fully differential exact results for the virtual photon correction
to single hard photon bremsstrahlung obtained using independent calculations,
both for $e^+e^-$ annihilation at high-energy colliders and for radiative return
applications.  The results are compared using Monte Carlo evaluations of the
matrix elements as well as by direct analytical evaluation of certain critical
limits. Special attention is given to the issues of numerical stability and
the treatment of finite-mass corrections.  It is found that agreement on the
order of $10^{-5}$ or better is obtained over most of the range of hard photon
energies, at CMS energies relevant to both high energy collisions and radiative
return experiments. 
\end{abstract}
\maketitle
%=======================================================================
\renewcommand{\baselinestretch}{1.1}
\section{\bf Introduction}
\label{intro}

The single hard bremsstrahlung process is critical in precision 
studies of Standard Model physics over a wide range of scales, from
the 1 GeV regime, where it is exploited in radiative return studies
of intermediate-energy hadron physics, to the 
100 GeV regime and beyond, where it is used for radiative return studies of the 
$Z$ resonance. Construction of a future linear collider would
push the energy range of interest toward the TeV regime. Precision studies of
per mille level effects in any of these regimes requires exact calculations
of all QED effects at order $\alpha^2$.  For radiative return, this means
that the virtual photon corrections to single hard photon radiation must
be included. 

Exact results for the  virtual corrections to single hard 
bremsstrahlung have been studied
by a number of groups.\cite{berends,in:1987,jmwy,rad6}
Comparisons of these results are important 
to be certain of their correctness at a high level of precision. 
In \Ref{jmwy}, we  have previously
presented comparisons of the results in \Refs{berends,in:1987,jmwy},
and in general a very good agreement was found. However,
it is clear from the comparisons in \Ref{jmwy} that small differences
appear at the level of the NNLL (next-to-next-to-leading-log), suggesting
differences in the levels of ``exactness'' in the calculations. 

These differences are largely due to how differential the results are,
and how the mass corrections are included. 
Both \Ref{jmwy} (JMWY) and \Ref{berends} 
include mass corrections, but
the former is fully differential, while the latter averages over the photon 
angle. The results of \Ref{in:1987} are fully differential,
but the mass corrections are neglected. These comparisons therefore
are not a complete test of the results in \Ref{jmwy} at the NNLL
level.

Another calculation of the exact virtual corrections to single hard 
bremsstrahlung has appeared recently.  An exact
differential result including mass corrections needed for an accurate
calculation of collinear emission is presented in \Ref{rad6} (KR).  
This result was obtained for use in
radiative return studies of hadronic final states.\cite{rad1,rad2,rad3} 
The virtual correction without explicit mass terms\cite{rad5} was obtained
for use in tagged photon experiments, and this was extended to the 
small angle regime for untagged experiments with the inclusion of finite mass 
corrections\cite{rad6}, which become significant in collinear emission,
even in very high-energy collisions. 

The results of KR have been implemented in the PHOKHARA 
MC~\cite{rad4,rad5,rad6,hans}, while the results of \Ref{jmwy} have been 
implemented in the \KKMC~\cite{kkmc:2001}.
In \Ref{hans2}, an indirect comparison of the
JMWY and KR results has been reported via comparison of the 
PHOKHARA and \KKMC's.  Agreement at the per mille level was reported
for selected observables. Further comparisons of these MC's were reported
in \Ref{jadcomp}, where the effect of including YFS (Yennie-Frautschi-Suura)
exponentiation\cite{yfs} in 
the \KKMC\ versus the unexponentiated approach of PHOKHARA plays an
important role in the differences found.

Presently, our interest is not in comparing the MC's but in comparing the
order $\alpha^2$ matrix elements for hard photon emission used in each.
A direct comparison of the matrix elements JMWY and KR  was reported in 
\Ref{compare1}, using a common MC evaluation of both expressions.  It 
was found that the expressions agree to $1.5\times 10^{-5}$ units of the 
Born cross section for a photon energy below 95\% of the beam energy at  
200 GeV CMS energy.

This level of agreement was the result of careful analytical work to 
cancel large terms in the expressions of KR. The published matrix elements
of KR and JMWY have very different forms, which is evident in part from
the appearance of mass corrections proportional to $(p_i\cdot k)^{-2}$ and
$(p_i\cdot k)^{-3}$ in the expressions of KR, where $p_i$ is an incoming 
fermion momentum and $k$ is the emitted hard photon momentum. Such terms are
absent in the JMWY expression, and we have verified that they cancel 
exactly in the KR expression as well, although this is not immediately obvious.
This cancellation should be implemented analytically to obtain a stable MC 
comparison of the results. Such an analytical cancellation is presented here.

The functions used to construct the stable comparisons were introduced in
\Ref{paris}.  A proof that the massless limits of the KR and JMWY
expressions agree at NLL (next-to-leading-log ) 
order was sketched and the size of
the difference in the mass corrections was calculated in \Ref{beijing}.
MC results for a comparison of the matrix elements in the 1 GeV radiative
return regime were also obtained.\cite{epiphany}  In this paper, we
will present details of the stabilization procedure needed for very
high precision comparisons, and present new MC comparisons which use
these versions of the matrix elements, which have been improved since the
publication of \Refs{compare1,paris,beijing,epiphany} for better 
behavior at lower CMS energies. 

An important part of this comparison is an examination of the 
finite-mass corrections used in the virtual corrections of KR and JMWY. 
The electron mass terms containing denominators of the form $p_i\cdot k$
give contributions of the same order as some of the ``massless'' 
contributions upon integration.  Therefore, the finite mass corrections of
both calculations are compared explicitly, and it is shown that both 
mass corrections are identical to next-to-leading-log (NLL) order. This 
supplements the result that the massless results agree to NLL order,
showing that the complete virtual corrections agree up to NNLL terms.
These NNLL terms are the subject of the numerical comparisons, both
with and without mass terms.

We need to point out that both approaches to the exact mass corrections
discussed here, that used by JMWY and that used by KR, should have the property
that, upon integrating over the final phase space, the resulting integrated
cross section should contain all finite terms that do not vanish 
for $m_e^2/s\rightarrow 0$. Both approaches use expansions therefore
in the parameter $m_e^2/s$. The JMWY approach uses the expansion constructed
by \Ref{berends1}. The KR approach uses the expansion of the respective
leptonic tensor coefficients through terms of order $m_e^4s/(2k\cdot p_i
)^3$, where $k$ is the radiated real photon four momentum and 
$p_i$ is an incoming $e^-$ or $e^+$ four momentum. It is therefore important to
know the relationship between these expansions both numerically and 
theoretically. We address this relationship matter in what follows.

The structure of this paper is as follows. In section \ref{analytical}, the
matrix elements of JMWY and KR are displayed explicitly in a compatible
format appropriate for comparisons, without the mass corrections in each
case. Both forms of the matrix elements are shown
in a form explicitly stable in collinear limits, which is suitable both
for MC evaluation at very high precision, and for an analytical investigation
of the collinear limits. A comparison of the collinear limits is 
presented in section \ref{NLL}, demonstrating that both matrix elements
agree to NLL order when the explicit mass terms are omitted.  Section
\ref{mass} examines the mass terms which are added in each case, and 
verifies that these both agree as well to NLL order, and to NNLL order
in the soft limit. A numerical comparison of the matrix elements is 
presented in section \ref{MC}, both at 1 GeV and 500 GeV CMS energies.
Finally, section \ref{conclusions} summarizes our conclusions, and the 
Appendix presents details on the special functions introduced to stabilize 
the collinear limits. 

\section{Virtual Corrections to Initial State Radiation}
\label{analytical}

The process we consider is initial-state radiation (ISR) in 
$e^+ e^-\rightarrow \mu^+\mu^-$, where one real and one virtual photon
are radiated on the electron-positron line. 
In \Ref{jmwy}, the matrix element was evaluated using helicity amplitude 
techniques of CALKUL\cite{berklei}, Xu {\it et al}\cite{xuzhangchang} 
and Kleiss-Stirling~\cite{KS}, together with
algebraic simplification using FORM~\cite{form}.
The mass corrections were then added following the methods in
\Ref{berends1}, after checking that the exact expression
for the mass corrections differs from the result obtained by the
latter methods by terms which vanish as $m_e^2/s\rightarrow 0$ in the
integrated cross section,
where $m_e$ is the electron mass and $s$ is the squared CMS energy. 

In \Ref{rad6}, the same ISR matrix
element's interference with the Born process is calculated
in terms of a leptonic tensor in such a way that all mass effects 
relevant in the collinear limits are obtained using an expansion
in powers of $m_e^2/p_i\cdot k$.  Thus, comparison 
of the two sets of results gives important
information on the two different methods of calculation and on the
two different treatments of the mass corrections.

The JMWY result used in these comparisons will be evaluated without $Z$ boson
exchange, to match the calculation of KR.  We denote the four momenta and 
helicity of the $e^-$, $e^+$, $f$, and $\bar f$ as $p_j$ and 
$\lambda_j$,\ $j=1,...,4$, respectively.  The tree-level cross section 
for single hard ISR for QED alone may be written as
\begin{equation}
\label{dsISR10}
{d\sISR{1}{0}\over d\Omega dr_1 dr_2} = {1\over 16 (4\pi)^4}
 \sum_{\lambda_i, \sigma} \left| \MISR{1}{0}\right|^2 ,
\end{equation}
where the squared matrix element, summed over helicities, is 
\begin{equation}
\label{MISR}
\sum_{\lambda_i, \sigma} \left|\MISR{1}{0}\right|^2 
= {16 e^6\over s^2 s' r_1 r_2} 
\left\{\left(t_1^2 + u_1^2\right) \left(1 - {2m_e^2 r_1\over s' r_2}\right)
     + \left(t_2^2 + u_2^2\right) \left(1 - {2m_e^2 r_2\over s' r_1}\right)
\right\} ,
\end{equation}
with $s = (p_1+p_2)^2$, $s' = (p_3 + p_4)^2$, 
$t_1 = (p_1 - p_3)^2$, $t_2 = (p_2 - p_4)^2$, 
$u_1 = (p_1 - p_4)^2$, $u_2 = (p_2 - p_3)^2$, $r_i = 2p_i\cdot k/s$, and
the solid angle $d\Omega$ refers to the direction of $p_3 - p_4$ relative
to $p_1 - p_2$ in the CM frame of the outgoing fermions.
The explicit mass corrections in \Eq{MISR} are those obtained using
the method of \Ref{berends1}. 

The same tree-level matrix element may be expressed in terms of the leptonic
tensor $L^{\mu\nu}$ of \Refs{rad5,rad6}. In the same notation, the 
leading-order contribution to the leptonic tensor may be written (in the
normalization of \Ref{rad5})
\begin{equation}
\label{L0}
L_0^{\mu\nu} = {e^4\over s(s')^2 r_1 r_2}\left\{ a_{00}^{(0)} s\eta^{\mu\nu}
  + a_{11}^{(0)} p_1^\mu p_1^\nu + a_{22}^{(0)} p_2^\mu p_2^\nu
  + a_{12}^{(0)} (p_1^\mu p_2^\nu + p_2^\mu p_1^\nu)\right\}
\end{equation}
with 
\begin{eqnarray}
\label{adef0}
a_{00}^{(0)} &=& -(1 - r_1)^2 - (1 - r_2)^2 + 
	{2 m_e^2 zv^2\over s r_1 r_2} ,\nonumber\\
a_{11}^{(0)} &=& -4z +  {8m_e^2 r_1\over s r_2},
\quad a_{22}^{(0)} = -4z + {8m_e^2 r_2\over s r_1},
\quad a_{12}^{(0)} = {-8m_e^2\over s},
\end{eqnarray}
where $z = s'/s = 1 - v$ with $v = r_1 + r_2$. 
This may be contracted with a final state tensor for a muon line, 
\begin{equation}
\label{M}
M^{\mu\nu} = e^2 \left\{p_3^\mu p_4^\nu + p_4^\mu p_3^\nu - (p_3\cdot p_4)
	\eta^{\mu\nu}\right\}\, ,
\end{equation}
to obtain the same squared matrix element as in \Eq{MISR}, 
\begin{eqnarray}
\label{kunsum0}
\sum_{\lambda_i, \sigma}
\left|\MISR{1}{0}\right|^2 &=& 16 L^{(0)}_{\mu\nu} M^{\mu\nu}\nonumber\\
&= & {8e^6\over s(s')^2 r_1 r_2}\left\{ 
	-ss'(2 a_{00}^{(0)} + a_{12}^{(0)})
 	+ a_{11}^{(0)}t_1 u_1 + a_{22}^{(0)} t_2 u_2 
	+ a_{12}^{(0)} (t_1 t_2 + u_1 u_2)\right\},\nonumber\\
\end{eqnarray}
where the coefficient is chosen to match the normalization of \Eq{MISR}, 
and explicit mass terms are kept only when enhanced by collinear factors. 
The mass terms in \Eq{MISR} and \Eq{kunsum0} agree in the collinear limits 
where $r_i$ is of order $m_e^2/s$. 

The initial-state virtual photon corrections to the cross section
may be expressed as 
\begin{equation}
\label{dsISR11}
{d\sISR{1}{1}\over d\Omega dr_1 dr_2} = {1\over 16 (4\pi)^4}
 \sum_{\lambda_i, \sigma} 2\,{\rm Re} \left[ (\MISR{1}{0})^* \MISR{1}{1}\right] , 
\end{equation}
where the matrix element for hard photon initial-state 
emission with one virtual photon may be expressed as 
\begin{equation}
\label{vdef1}
\MISR{1}{1} = {\alpha\over 4\pi} \MISR{1}{0} (f_0 + f_1 I_1 + f_2 I_2),
\end{equation}
where $f_i$ are scalar form factors and $I_i$ are spinor factors defined in
\Ref{jmwy}:
\begin{subequations}
\begin{eqnarray}
I_1 &=& \sigma \lambda_3 s_{\lambda_1}(p_1,k)
s_{-\lambda_1}(p_2,k)\nonumber\\\nopagebreak
& &\quad\times
{s_{\lambda_3}(p_4,p_2)s_{-\lambda_3}(p_2,p_3)
- s_{\lambda_3}(p_4,p_1)s_{-\lambda_3}(p_1,p_3)
\over s_{-\sigma}(p_1,p_2) s_{-\sigma}(p_3,p_4)
s^2_{\sigma}(p_{21},p_{34})} ,
\\
I_2 &=& \lambda_1\lambda_3{s_{\lambda_1}(p_1,k) s_{-\lambda_1}(p_2,k)
s_{\lambda_3}(p_4,k) s_{-\lambda_3}(p_3,k) \over
s_{-\sigma}(p_1,p_2) s_{-\sigma}(p_3,p_4)
s^2_{\sigma}({\tilde p}_{12},p_{34})} ,
\end{eqnarray}
\end{subequations}
where $\lambda_i$ is the helicity of the fermion with momentum $p_i$ and 
$\sigma$ is the photon helicity. The spinor product is
$s_{\lambda}(p, q) = {\bar u}_{-\lambda}(p) u_{\lambda}(q)$, and
\begin{equation}
(p_{ij}, {\tilde p}_{ij}) = \left\lbrace {(p_i, p_j)\atop(p_j,p_i)} \right.
	\hbox{\ for\ } \sigma = \left\lbrace {\lambda_i\atop\lambda_j}\right.\ .
\end{equation}
The form factors $f_i$ are given (without mass corrections) by 
\begin{subequations}
\begin{eqnarray}
f_0 &=& 4\pi \BYFS(s) + 2\left(L - 1 - i\pi\right)
+ {r_2\over 1-r_2}
+ {r_2(2+r_1)\over(1-r_1)(1-r_2)} \left\{\ln\left({r_2\over z}\right)
+ i\pi\right\}
\nonumber\\
&-& \left\{3v + {2r_2 \over 1-r_2}\right\} \Lf_1\left(-v\right)
+ {v \over (1-r_2)}\; R_1(r_1,r_2) + r_2 R_1(r_2,r_1),
\\
f_1 &=& {r_1 - r_2\over 2(1-r_1)(1-r_2)}
+ {z\over (1-r_1)(1-r_2)}\left(r_2 + {1+z\over 2(1-r_1)}\right)
        \left\{\ln\left(r_2\over z\right) + i\pi\right\}
\nonumber\\
&+& {z\over 1-r_2}\left\{{1\over2} R_1(r_1,r_2) + r_2 R_2(r_1, r_2)\right\}
+ {v\over 4}\left\{R_1(r_1,r_2)\delta_{\sigma,1} +
R_1(r_2,r_1)\delta_{\sigma,-1}\right\},
\\
f_2 &=& v + {r_1 r_2 - v/2\over (1-r_1)(1-r_2)} - 2vz\Lf_3(-v)\nonumber\\
&+& {z\over (1-r_1)(1-r_2)} \left(2 - r_2 + {r_2-r_1\over 2(1-r_1)}\right)
	\left\{\ln\left(r_2\over z\right) + i\pi\right\}
\nonumber\\
&+& {z\over 1-r_2}\left\{{1\over2} R_1(r_1,r_2) + (2-r_2) R_2(r_1, r_2)\right\}
\nonumber\\
&+& {r_1-r_2\over 4}\left\{R_1(r_1,r_2)\delta_{\sigma,1}
+ R_1(r_2,r_1)\delta_{\sigma,-1}\right\}
\end{eqnarray}
\end{subequations}
for photon helicity $\sigma = \lambda_1$.  When $\sigma =
-\lambda_1$, $r_1$ and $r_2$ must be interchanged. 
Here, $L = \ln(s/m_e^2)$ is the ``large logarithm'' which may be used
 in a leading log expansion of the results. Thus, leading log (LL) refers
to order $L^2$, NLL to order $L$, and NNLL to order $L^0 = 1$ in this
expansion.

The infrared divergence is
contained in the virtual YFS form factor defined by \cite{yfs}
\begin{equation}
4\pi \BYFS(s) = \left(2\ln{m_\gamma^2\over m_e^2} + 1\right)
	\left(L - 1 - i\pi\right) 
	- L^2 - 1 + {4\pi^2\over3}
	+ i\pi\left(L - 1\right).
\end{equation}
We also make use of functions
\begin{subequations}
\begin{eqnarray}
R_1(x, y) &=& \Lf_1(-x)\left\{\ln\left({1-x\over y^2}\right) - 2\pi i
\right\}
\nonumber\\
 &+& {2(1-x-y)\over 1-x}\; \Sf_1\left({y\over 1-x}, {x(1-x-y)\over 1-x}
\right),
\\
R_2(x, y) &=& \Lf_1(-x) - {1\over2}\Lf_1^2(-x) + {x\over x+y} \Lf_2(-x)
	+ {y\over (x+y)(1-x-y)} \Lf_2\left(y\over 1-x-y\right)
\nonumber\\
&+& \Lf_2(-x)(\ln y + i\pi) + \left({1-x-y\over 1-x}\right)^2\ \Sf_2
	\left({y\over 1-x}, {x(1-x-y)\over 1-x}\right),
\end{eqnarray}
\end{subequations}
with $\Lf_n(x)$, $\Sf_n(x, y)$ defined in \Eq{difdef} of the Appendix.

The functions $\Lf_n$ and 
$\Sf_n$ play an important role in canceling large factors in differences 
between logarithms or dilogarithms with slightly different arguments. 
Therefore, although the expressions
for $f_0, f_1, f_2$ are analytically equivalent to those of \Ref{jmwy},
they are more stable when evaluated numerically. The functions $R_1$ and 
$R_2$ are related to the function $R$ used in \Ref{jmwy} by 
\begin{eqnarray}
R_1(x,y) &=& {1\over x} R(x,y), \nonumber\\
R_2(x,y) &=& {1\over x}\left\{{1\over2}R_1(x,y) - \Lf_1(-x) + \ln\left({
	y\over 1-x-y}\right) + i\pi\right\}.
\end{eqnarray}
The functions $R_i$ are defined so 
that both are finite for $x, y \rightarrow 0$
with $x, y$ positive. This implies that  $f_1$ and $f_2$ are finite for 
$r_1, r_2 \rightarrow 0$.  The spinor products $I_1$ and $I_2$ vanish in
this limit, since $|s_\lambda(p_i,k)|^2 = sr_i$. Therefore, the $f_1$ and
$f_2$ terms are absent in the collinear limits.

The virtual corrections $a_{ij}^{(1)}$ to the coefficients 
$a_{ij}$ are calculated without explicit
mass terms in \Ref{rad5}. The infrared divergence in that result is canceled
by adding the contribution of an additional soft real photon with energy 
below a cutoff $v_{\min}$ as a fraction of the beam energy.
To facilitate comparisons, the additional soft photon contribution will
be removed, so that the pure virtual corrections are compared in
each case. This requires decomposing the infrared-divergent term
\begin{equation}
\label{origIR}
a_{ij}^{\rm IR} = a_{ij}^{(0)}\left\{ 2(L-1)\ln v_{\min} 
 + {3\over2} (L + \ln z) - 2 + {\pi^2\over 3}\right\}
\end{equation}
in \Ref{rad5} into a virtual contribution $a_{ij}^{(0)} F_{\rm IR}$ and
a contribution due to an additional real soft photon.  The correct
decomposition is found to be 
\begin{equation}
\label{IR-decomposition}
a^{\rm IR}_{ij} = a_{ij}^{(0)}(F_{\rm IR} + 2\pi {\widetilde B}_{\rm YFS}(s, v_{\min}))
\end{equation}
with 
\begin{equation}\label{FIR}
F_{\rm IR} =
 2\pi\Re\;\BYFS(s) + L - 1 + {3\over2}\ln z 
\end{equation}
and the real YFS soft photon form factor 
\begin{equation}
2\pi {\widetilde B}_{\rm YFS}(s, v_{\min}) 
     = (L-1)\left\{\ln{m_e^2\over m_\gamma^2}
	+ 2\ln v_{\min}\right\} + {1\over2} L^2 - {\pi^2\over3}.
\end{equation}
This can be verified by comparing the soft limit of the virtual correction to
the known result, as is done at the end of this section.

We will use the functions $\Lf_n$ and $\Sf_n$ defined in the 
Appendix to stabilize the
results and clarify the collinear behavior. Then $a_{ij} = a_{ij}^{(0)}
+ {\alpha\over\pi} a_{ij}^{(1)}$, with 
\begin{equation}
a_{ij}^{(1)} = {\alpha\over\pi} \left(a_{ij}^{(0)} F_{\rm IR} + 
    a_{ij}^{(1,0)} +  a_{ij}^{(1,m)}\right),
\end{equation}
where the 
massless parts $a_{ij}^{(1,0)}$ of the non-IR-divergent part of the 
virtual corrections are given by 
\begin{subequations}
\begin{eqnarray}
a_{00}^{(1,0)} &=& {vz\over4} - {r_1 r_2\over 2} 
	+ {z\over 2} \ln z + r_1 r_2\ \Lf_1(-v)
+ {r_2\over 2} \left\{ 4 - r_2 - {3(1+z)\over 1-r_1}\right\}
	\ln\left({r_2\over z}\right)\nonumber\\
&+& \left\{r_1 + {r_2\over z}\left[1 + (1-r_2)^2\right] \right\}
	\left\{ z\ln\left({r_2\over z}\right)\ \Lf_1(-r_1)
	- \Sf_1\left(-{r_1\over z}, -{r_2\over z}\right) 
	\right\}\nonumber\\
&+& (r_1 \leftrightarrow r_2),
\end{eqnarray}
\begin{eqnarray}
a_{11}^{(1,0)} &=& 3 + z + {(1+z)^2\over 1 - r_1} - 6\ln z - 4\;\Lf_1(-v)
\nonumber\\ 
&-& 4(1 + r_1 - r_1 r_2)\left\{\Lf_1(-v) + \Lf_2(-v)\right\}
+ 2\left({v\over z} - z(1-r_2)\right)\Lf_1\left({r_1 r_2\over z}\right)
\nonumber\\
&+& 2(1+z)\left\{(1-r_2)\;\Lf_1(-r_2) - \Lf_1\left({-r_1\over 1-r_2}\right)
	- {v\over z}\ \Lf_1\left({r_2\over z}\right)\right\}
\nonumber\\
&+& 2\;\Lf_1(-r_1) - {2\over z}\ \Lf_1\left({r_1\over z}\right)
+ 2r_2 z \ln\left({r_1\over z}\right) \left\{\Lf_1(-r_2) + \Lf_3(-r_2)\right\}
\nonumber\\
&+& r_1 z\ln\left({r_2\over z}\right) \left\{ 
   r_1\left({1-r_2\over 1-r_1}\right)^2 + 4\;\Lf_1(-r_1) 
   + 2(1-r_2)^2\;\Lf_3(-r_1)\right\}
\nonumber\\
&-& 2r_2\;\Sf_1\left({-r_1\over z}, {-r_2\over z}\right)
+ {2\over z}\ \Sf_2\left({-r_1\over z}, {-r_2\over z}\right)
\nonumber\\
&-& 2(2 + r_1 - 2 r_2)\Sf_1\left({-r_2\over z}, {-r_1\over z}\right)
+2z\;\Sf_2\left({-r_2\over z}, {-r_1\over z}\right),
\\
a_{22}^{(1,0)} &=& a_{11}^{(1,0)}(r_1 \leftrightarrow r_2), 
\end{eqnarray}
\begin{eqnarray}
a_{12}^{(1,0)} &=& {z(1+z)\over 2(1-r_1)(1-r_2)} + 2r_1 r_2 - {v\over 2}
\nonumber\\
&-& z\;\Lf_1(-v) - 2z(1-r_1 r_2)\ \Lf_2(-v) 
	- z\;\Lf_1\left({r_1r_2\over z}\right)
\nonumber\\
&+&2z(r_1 r_2 + r_2-1)\left\{{1\over 1-r_1} + \Lf_2(-r_1)\right\}
	\ln\left({r_2\over z}\right)
\nonumber\\
&+& {z^2\over (1-r_2)^2}\ln\left({r_2\over z}\right)
- 2(1-r_1)\ \Sf_1\left({-r_1\over z}, {-r_2\over z}\right)
+ 2\;\Sf_2\left({-r_1\over z}, {-r_2\over z}\right)
\nonumber\\
&+& (r_1 \leftrightarrow r_2).
\end{eqnarray}
\end{subequations}
These expressions are exactly equal to those of \Ref{rad5}, but 
written in terms of the functions $\Lf_n$ and $\Sf_n$.

A comparison of the collinear limits of the two matrix elements will 
be presented in the following section.
A numerical study comparing these two forms of the virtual photon
correction to hard bremsstrahlung will be presented in section
\ref{MC}, together with a comparison of the mass corrections added
to obtain the complete matrix elements in each case.

\section{NLL Comparision of the Matrix Elements}
\label{NLL}

In this section,  we will verify that the expressions of \Ref{jmwy} and 
\Ref{rad5} agree analytically to NLL order, meaning that they agree
through order $L$ when integrated over $r_i$.  For this purpose, we
may examine both expressions in the limit of collinear emission, which
is sufficient to determine the NLL result.  We will
consider the particular limit where $r_1$ becomes small, so that the photon is 
collinear with the incoming electron. 

In the collinear limits, the virtual correction to the squared matrix 
element is proportional to the tree-level result,
\begin{equation}
\label{rvsum}
\sum_{\lambda_i, \sigma} 2\;{\Re}\;\left({\MISR{1}{0}}^* \MISR{1}{1}\right) =
{\alpha\over 2\pi}\;{\Re}\;\langle f_0^{\rm NLL}\rangle 
    \sum_{\lambda_i, \sigma} \left|\MISR{1}{0}\right|^2 ,
\end{equation}
where $\langle f_0^{\rm NLL} \rangle$ is the average of $f_0$ over helicities,
evaluated for small $r_1$ or $r_2$ small, \cite{jmwy}
\begin{eqnarray}
\label{NLL-result}
{\Re}\;\langle f_0^{\rm NLL}\rangle &=& 2\pi{\Re} B_{\rm YFS}(s) + L - 1 
	+ 3\ln (1-r_1) + 2\ln r_1 \ln (1-r_2) 
\nonumber\\
&-& \ln^2(1-r_1) + 2\Sp(r_1) + {r_1(1-r_1)\over 1 + (1-r_1)^2}  
+ (r_1 \rightarrow r_2).
\end{eqnarray}

Since the $f_1, f_2$ terms in \Eq{vdef1} both vanish in collinear limits, they
do not contribute to NLL order.  In the limit when $r_1 \rightarrow 0$, the 
NLL expression may be written in the form
\begin{equation}
\label{collinear}
{\Re}\;\langle f_0^{\rm NLL} (r_1\rightarrow 0)\rangle
= F_{\rm IR} + 2\ln z \ln r_1
- \ln^2 z + 2\ \Sp(v) + {vz\over 1 + z^2}.
\end{equation}

The result of \Ref{rad5} can be compared to this expression using 
\Eq{kunsum0} with $a_{ij}^{(0)}$ replaced by $a_{ij}^{(1)}$ in the
collinear limit. The virtual correction to the squared 
matrix element may be written
\begin{equation}
  \sum_{\lambda_i, \sigma} 2\;{\Re} \left({\MISR{1}{0}}^* 
\MISR{1}{1}\right) = 16L^{(1)}_{\mu\nu}M^{\mu\nu}
= {\alpha\over 2\pi} {\Bigg\{} 2F_{\rm IR}
\sum_{\lambda_i, \sigma}\left|\MISR{1}{0}\right|^2  \nonumber
\end{equation}
\begin{equation}
\label{kunsum1}
+ {8e^6\over s(s')^2 r_1 r_2} \left[
        - ss'\left(2a_{00}^{(1,0)} + a_{12}^{(1,0)}\right)
        + a_{11}^{(1,0)}t_1 u_1 + a_{22}^{(1,0)} t_2 u_2
        +  a_{12}^{(1,0)} (t_1 t_2 + u_1 u_2)\right]{\Bigg\}}.
\end{equation}

In the collinear limit, this expression should again be proportional to the 
tree level result, \Eq{MISR} without the mass terms.  When $r_1$ is small, 
$t_2 = z t_1 + {\cal O}(r_1)$ and $u_2 = z u_1 + {\cal O}(r_1)$, 
allowing  \Eq{kunsum1} to be written as 
\begin{equation}
\sum_{\lambda_i, \sigma} 2\;{\Re}\;\left({\MISR{1}{0}}^* \MISR{1}{1}\right) =
{\alpha\over 2\pi} \left\{ f' \sum_{\lambda_i, \sigma}
	\left|\MISR{1}{0}\right|^2
 + {4e^6\Delta\over szr_1 r_2} \right\}
\end{equation}
with 
\begin{eqnarray}
f' &=& 2F_{\rm IR}
 + {1\over a_{00}^{(0)}} \left(a_{00}^{(1,0)} + {\Delta\over 4}\right),
\nonumber\\
\Delta &=& {a_{11}^{(1,0)}\over z} + za_{22}^{(1,0)} - 2a_{12}^{(1,0)} 
	- 4a_{00}^{(1,0)},
\end{eqnarray}
where $a_{00}^{(0)} = - (1 + z^2)$ is the massless limit of the tree-level
coefficient.  When $r_2$ is small, the same form is obtained with $a_{11}$
and $a_{22}$ interchanged.

We may check that $\Delta = 0$ and $f' = {\Re}\;f_0$ 
in the collinear limits. 
Evaluating the coefficient functions in the limit when $k$ is collinear 
with $p_1$ gives
\begin{subequations}
\begin{eqnarray}
a_{00}^{(1,0)} (r_1 \rightarrow 0) &=& -{vz\over2} + (1 + z^2) \left\{ 
 {1\over 2} \ln^2 z - \ln z \ln r_1 - \Sp(v)\right\},\\
a_{11}^{(1,0)} (r_1 \rightarrow 0) &=& 3 + 2z
- 2\ln r_1 \left\{ z\ln z +\ \Lf_1(-v) +\ \Lf_2(-v)  - {1\over2}(1 + v)\right\}
\nonumber\\
&+& \ln z\left\{\Lf_1(-v) + 2\;\Lf_2(-v) - 8\right\}
+ z\ln^2 z - 2z\Sp(v) \nonumber\\
&-& 11\;\Lf_1(-v) - \Lf_1^2(-v) 
- 5\;\Lf_2(-v) + 2\;\Sf_1(0,v) - 2\;\Sf_2(0,v), \\
a_{22}^{(1,0)} (r_1 \rightarrow 0) &=& {1\over z} + 2z +
\ln r_1 \left\{{1\over z} - 4z\ln z - 2\;\Lf_1(-v) - 2\;\Lf_2(-v)\right\}
\nonumber\\
&+& \ln z \left\{ \Lf_1(-v) + 2\;\Lf_2(-v) - {3\over z} -4\right\} + 2z\ln^2 z
- \Lf_1(-v) \left({2\over z} + 7\right) 
\nonumber\\
&-& 4z\Sp(v) -\ \Lf_1^2(-v) - 5\;\Lf_2(-v) + 2\;\Sf_1(0,v) - 2\;\Sf_2(0,v),
\\
a_{12}^{(1,0)} (r_1 \rightarrow 0) &=& 3 + z + 
\ln r_1\left\{1 - 2\;\Lf_1(-v) - 2\;\Lf_2(-v)\right\} \nonumber\\
&+& \ln z \left\{ \Lf_1(-v) + 2\;\Lf_2(-v) - 6 - 2z \right\} - 10\;\Lf_1(-v)
\nonumber\\
&-& \Lf_1^2(-v) - 5\;\Lf_2(-v) + 2\;\Sf_1(0,v) - 2\;\Sf_2(0,v).
\end{eqnarray}
\end{subequations}
Combining these relations confirms that $\Delta = 0$ and 
$f' = {\Re}\;f_0$, so the results have the same collinear limits without
explicit mass terms.  This implies that the results without mass terms 
agree to NLL order.  

Since the collinear limits agree, the soft limits also agree.  
Specifically, for $v \rightarrow 0$, 
\begin{equation}
a_{00}^{(1,0)} = 0, \quad a_{11}^{(1,0)} = a_{22}^{(1,0)} = 
	a_{12}^{(1,0)} = -3,
\end{equation}
and $f' = {\Re}\;f_0 = 4\pi\; {\rm Re} \BYFS(s) + 2(L-1)$. This verifies
that the virtual part of the infrared contribution was identified 
correctly in \Eq{IR-decomposition}.

\section{Comparison of Mass Corrections} 
\label{mass}

Explicit mass corrections are added in the JMWY result via the prescription 
of \Ref{berends1} and
sec.\ 3 of \Ref{jmwy1}, leading to a correction which may be 
expressed as 
\begin{equation}
\label{jmwymas}
\sum_{\lambda_i, \sigma} 2\;{\Re}\;\left({\MISR{1}{0}}^* \MISR{1}{0}\right)_m 
 = - {m_e^2\over s'} {\alpha\over \pi} \fNR(s') {16 e^6\over s'}
\left\{  {t_2^2 + u_2^2\over s^2 r_1^2} + {t_1^2 + u_1^2\over s^2 r_2^2} 
	\right\},
\end{equation}
where the non-radiative virtual correction factor is 
\begin{equation}
\label{fNR}
\fNR(s') = 4\pi \BYFS(s') + 2\ln\left({s'\over m_e^2}\right) - 2 
\end{equation}
with 
\begin{equation}
4\pi\BYFS(s') = 4\pi\BYFS(s) + \ln z\left(2\ln{m_\gamma^2\over m_e^2} - 2L
+ i\pi\right) - \ln^2 z.
\end{equation}
The first term in brackets in \Eq{jmwymas} is due to $e^+$ line emission,
and the second is due to $e^-$ line emission. 

This mass prescription is designed to produce the essential mass corrections
in the collinear limits when the photon is emitted along the electron or 
positron line. In these limits, $r_i = m_e^2 v/s$, where $i$ labels the 
momentum of the collinear incoming fermion line. 
The mass corrections of \Ref{rad6}
may be compared analytically in this limit.  

The explicit mass correction of KR is given by 
\begin{equation}
\label{kunmas}
\sum_{\lambda_i, \sigma} 2\;{\Re}\;\left({\MISR{1}{0}}^* \MISR{1}{0}\right)_m =
16 L_{\mu\nu}^{(1,m)} M^{\mu\nu},
\end{equation}
where only the mass corrections to the virtual part of the leptonic tensor
are included, so that it is calculated using the coefficient functions 
\begin{equation}
a_{ij}^{(m)} = {\alpha\over\pi}\left(a_{ij}^{(0,m)} F_{\rm IR} + a_{ij}^{(1,m)}
\right)
\end{equation}
with $F_{\rm IR}$ as in \Eq{FIR} and $a_{ij}^{(0,m)}$ given by the explicit 
mass terms in \Eq{adef0}.  The functions
$a_{ij}^{(1,m)}$ are given in \Ref{rad6}, and may be written in terms of 
$\rho_i = s r_i/m_e^2$ and the stabilized functions $\Lf_n$ and $\Sf_n$ as 
\begin{subequations}
\begin{eqnarray}
a_{00}^{(1,m)} &=& {zr_2\over \rho_1}\left\{-\ln z \left(
    2L - 4\ln \rho_1 + \ln z\right) + 4\Sp(v)\right\} - zr_2\Sf_1(1,-\rho_1)
\nonumber\\
 &-& {1\over2} zr_2 N_1\left(\rho_1, {1\over z} - 4\right) + vr_2 N_3(\rho_1) 
 + (r_1,\rho_1 \leftrightarrow r_2,\rho_2),
\\
a_{11}^{(1,m)} &=& {4r_1\over\rho_2} \left\{-\ln z \left(
    2L - 4\ln \rho_2 + \ln z\right) + 4\;\Sp(v)\right\} 
    - 4r_1\left({z\over v}+2\right)\ \Sf_1(1,-\rho_1)
\nonumber\\
    &+& {r_1\over v} \left\{4zN_3(\rho_2) + \left(2z - {1\over z}\right)
	N_1\left(\rho_2, {2(1 - 2z)^2\over 1-2z^2}\right)
        + 6 N_2(\rho_2) + 2N_4(\rho_2)\right\}
\nonumber\\
   &+& {zr_2\over v}\left\{4N_3(\rho_1) - N_1(\rho_1,0) + 2z\left(N_2(\rho_1)
    + N_4(\rho_1)\right)\right\},
\\
a_{22}^{(1,m)} &=& a_{11}^{(1,m)} (r_1, \rho_1 \leftrightarrow r_2, \rho_2),
\\
a_{12}^{(1,m)} &=& {zr_2\over v} 
    \left\{4N_3(\rho_1) + 4N_2(\rho_1) + 2N_4(\rho_1) + 2\;\Sf_1(1,-\rho_1)
     \right\} - r_2 N_1\left(\rho_1,{z\over v}\right) 
\nonumber\\
         &+& (r_1,\rho_1 \leftrightarrow r_2,\rho_2).
\end{eqnarray}
\end{subequations}
where the functions $N_i$ are defined by 
\begin{subequations}
\begin{eqnarray}
N_1(\rho,w) &=& \Lf_2(\rho-1) - w\;\Lf_2(\rho-1), \\
N_2(\rho) &=& \Sf_1(z,v-\rho) + \Lf_1(-v)\ \Lf_1\left({\rho\over v}-1\right)
\nonumber\\
&=& \Sf_1(v,\rho-v) - {1\over z}\ln\rho\  \Lf_1\left({v-\rho\over z}\right) 
\quad (\rho < 1),
\\
N_3(\rho) &=& {1\over \rho}\left(1 - \ln\rho - \Sf_1(1,-\rho)\right)
\nonumber\\
&=& -\ln\rho\ \Lf_2(-\rho) -\ \Sf_2(0,\rho) \quad (\rho < 1),\\
N_4(\rho) &=& \Sf_2(z,v-\rho) - {1\over v}\ \Lf_1(-v)\ \Lf_2\left({\rho\over v} 
    - 1\right) - {1\over vz}\ \Lf_1\left({\rho\over v} - 1\right)
\nonumber\\
&=& -\Sf_2(v,\rho-v) - {1\over z^2} \ln\rho\ \Lf_2\left({v-\rho\over z}\right)
\quad (\rho < 1).
\end{eqnarray}
\end{subequations}
All of the expressions $N_i$ are finite in the collinear limit 
$\rho\rightarrow v$: 
\begin{subequations}
\begin{eqnarray}
N_1(v, w) &=& {1\over z} -\ \Lf_1(-z)\left({1\over z} + w\right),\\
N_2(v) &=& \Lf_1(-v) +\ \Lf_1(-z) , \\
N_3(v) &=& -\ \Sf_2(0,v) - \ln v\ \Lf_2(-v) ,\\
N_4(v) &=&  -{1\over2}\left\{\Lf_2(-v) + {1\over z}\left(1 + 
         \Lf_1(-z)\right)\right\}.
\end{eqnarray}
\end{subequations}
Also, $\Sf_1(1,-\rho) \rightarrow \Sf_1(0,v) -\ln v\ \Lf_1(-v)$ in the 
collinear limit.  All of the functions $N_i$ vanish in the limit 
$\rho\rightarrow\infty$, where masses are completely neglected.

In the collinear limit $\rho_1 \rightarrow v$, corresponding to $e^+$ line
emission, the total mass correction
functions $a_{ij}^{(m)}$ (including the infrared part) simplify to 
\begin{subequations}
\begin{eqnarray}
a_{00}^{(m)} &=& {m_e^2 v\over sr_1}\left\{2z F_{\rm IR} 
    + (3z-1)(\ln v \ln z + \Sp(v)) - 2zL \ln z - {v^2\over z} \ln v \right.
\nonumber\\
&-&  \left. z\ln^2 z + {v\over 2}\right\},
\\
a_{11}^{(m)} &=& -{m_e^2 v\over sr_1}\left\{\ln v \left({v\over z} + 2z +
  4z\; \Lf_2(-v)\right) + 1 + z\right.
\nonumber\\
&-& \left. 2 z^2\; \Lf_1(-v) + z^2\; \Lf_2(-v) + 4z\;\Sf_2(0,v)\right\},
\\
a_{22}^{(m)} &=& {m_e^2 v\over sr_1}\left\{8F_{\rm IR}
-8L\ln z - {1\over z^3}(1 + z - 4z^2 - 8z^3)\ln v
  + 8\ln v \ln z + 8\Sp(v)\right.
\nonumber\\
&+& 4z\ln v\ \left(\Lf_1(-v) - \Lf_2(-v)\right) - 4\ln^2 z 
+ 6\;\Lf_1(-v) - \Lf_2(-v)
\nonumber\\
&-&\left. 4z\left(\Sf_1(0,v) + \Sf_2(0,v)\right)
   - {1\over z^2}(1 + z - 2z^2)\right\},
\\
a_{12}^{(m)} &=& {m_e^2 v\over sr_1}\left\{
2z\ln v\left(\Lf_1(-v) - 2\;\Lf_2(-v)\right)
- \left(4 + {v^2\over z^2}\right) \ln v \right.
\nonumber\\
&+& \left. 4z\;\Lf_1(-v) - z\;\Lf_2(-v) - 2z\;\Sf_1(0,v) - 4z\;\Sf_2(0,v)
 - {1\over z}
\right\}.
\end{eqnarray}
\end{subequations}
Note that there are no terms proportional to $m^4/r_1^2$, so that all
such factors in \Ref{rad6} cancel when the terms are combined into the
stabilized functions used here. 

To verify that this result agrees with the JMWY result to NLL order,
it is necessary to check all terms with explicit logarithms $L$ in 
the collinear limit. In the $e^+$ line emission case, where $k$ is 
collinear with $p_1$, 
\begin{equation}
\label{krsummas}
\sum_{\lambda_i, \sigma} 2\;{\Re}\;\left({\MISR{1}{0}}^* \MISR{1}{0}\right)_m =
{-4 e^6\over s^2(s')^2 r_1 r_2}\left\{{4\over z} (t_2^2 + u_2^2) {f'}_m
- z s^2 \Delta_m\right\}
\end{equation}
with 
\begin{eqnarray}
{f'}_m = {a_{11}^{(m)}\over 4z}
+ {z a_{22}^{(m)}\over 4} - {a_{12}^{(m)}\over 2} &=& 2F_{\ IR} - 2L\ln z + 
{\cal O}(L^0)\nonumber\\
&=& 2L\left(\ln{m^2_\gamma\over m^2_e} + 1 - \ln z\right) -  L^2 
	+ {\cal O}(L^0)
\end{eqnarray}
and 
\begin{equation}
\Delta_m = {a_{11}^{(m)}\over z} + z a_{22}^{(m)}
- 4a_{00}^{(m)} = {\cal O}(L^0).
\end{equation}
Therefore, 
\begin{equation}
\label{krmas}
\sum_{\lambda_i, \sigma} 2{\Re}\left({\MISR{1}{0}}^* \MISR{1}{0}\right)_m 
= - {m_e^2\over s'} {\alpha\over \pi} {16 e^6\over s'}
\left(  {t_2^2 + u_2^2\over s^2 r_1^2} \right)
\left\{2L\left(\ln{m_\gamma^2\over m_e^2} + 1 - \ln z\right) - L^2 \right\}
\end{equation}
to NLL order.  This agrees with the $e^+$ line emission part of \Eq{jmwymas} 
to the same order, since 
\begin{equation}
f_{\rm NR}(s') = 2L\ln{m_\gamma^2\over m_e^2} - L^2 + 2L(1 - \ln z) 
	+ {\cal O}(L^0).
\end{equation}
Therefore, the both expressions for the explicit mass terms agree analytically
through NLL order. 

We can also check that the mass terms agree exactly in the soft collinear
limit.  Taking $v \rightarrow 0$ in the collinear results for the 
coefficients $a_{ij}^{(m)}$ gives 
\begin{equation}
a_{00}^{(m)} = 2F_{\rm IR}, \quad a_{11}^{(m)} = -1/2, \quad
a_{22}^{(m)} = 8F_{\rm IR} + 3/2, \quad a_{12}^{(m)} = 1/2,
\end{equation}
which gives to \Eq{krsummas} with $f'_m = 2F_{\rm IR}$ and $\Delta_m = 0$.  
In the soft limit, $f_{\rm NR} = F_{\rm IR}$, so the mass corrections are 
identical in the soft collinear limit. 

\section{Monte Carlo Comparisons}
\label{MC}

Since both of the results compared are fully differential, they are
well suited to MC implementation.  The JMWY results were developed for
the \KKMC\ described in \Ref{kkmc:2001}, which is for high-energy fermion
pair production. The results from \Ref{rad6}
are implemented in the PHOKHARA MC developed for radiative return
applications at $\Phi$ and B factories. Structurally, these MC's are 
considerably different, with \KKMC\ implementing a YFS-exponentiated 
algorithm, and PHOKHARA unexponentiated. Some comparisons of the 
MC's have been reported.\cite{hans2, jadcomp} Here, we will not 
compare the MC's, since that would test not just the comparison of
the matrix elements, but also be affected by the presence or absence
of exponentiation. Instead, we will implement both virtual photon corrections
in the same MC, with the sole purpose of comparing the two matrix elements.

Since the results of JMWY were developed for use in the 
YFS-based Monte Carlo (MC) program \KKMC \cite{kkmc:2001}, we will look at how
the results of \KKMC\ would change if the virtual correction of KR are
substituted for those of JMWY in the same program.  We calculate the
YFS residuals $\bar\beta_1^{(i)}$ as defined in \Refs{kkmc:2001,compare1}.
The residuals of interest here are for single photon emission, 
$\beta_1^{(i)}$ for $i = 1, 2$
which were compared previously for several different virtual photon 
corrections in \Refs{compare1,paris,beijing,epiphany}.   The superscript
$(i)$ denotes the power of $\alpha$ relative to the born level, 
$\beta_0^{(0)} = d\sigma_{\rm Born}/d\Omega$.  Thus, $i = 1$ corresponds to 
tree-level hard photon emission, and $i=2$ corresponds to the virtual
corrections. The residuals are integrated
up to a cutoff $v_{\rm max}$ on the photon energy fraction to produce
cross-sections denoted $\sigma_{\beta_1}^{(i)}$.  

Results for comparisons of these YFS residuals were previously reported
in \Refs{jmwy,compare1,paris,beijing,epiphany} for the virtual photon 
corrections of \Refs{in:1987,berends} at a CMS energy of 200 GeV.  Some
results were also reported in \Ref{compare1} for 500 GeV. It was found 
in \Ref{compare1} that at 200 GeV, all of the fully 
differential results
(excluding \Ref{berends}, which is not fully differential) agreed to within
$5\times 10^{-5}$ units of the Born cross section across the full range
of $v$ investigated (up to  $v_{\rm max} = 0.9999$). Agreement to within
$2\times 10^{-4}$ units of the Born cross section was found at 500 GeV.  

\begin{figure*}
\label{fig1}
\setlength{\unitlength}{1in}
\begin{picture}(6.5,3.0)
\put(0,0){\includegraphics[width=2.75in]{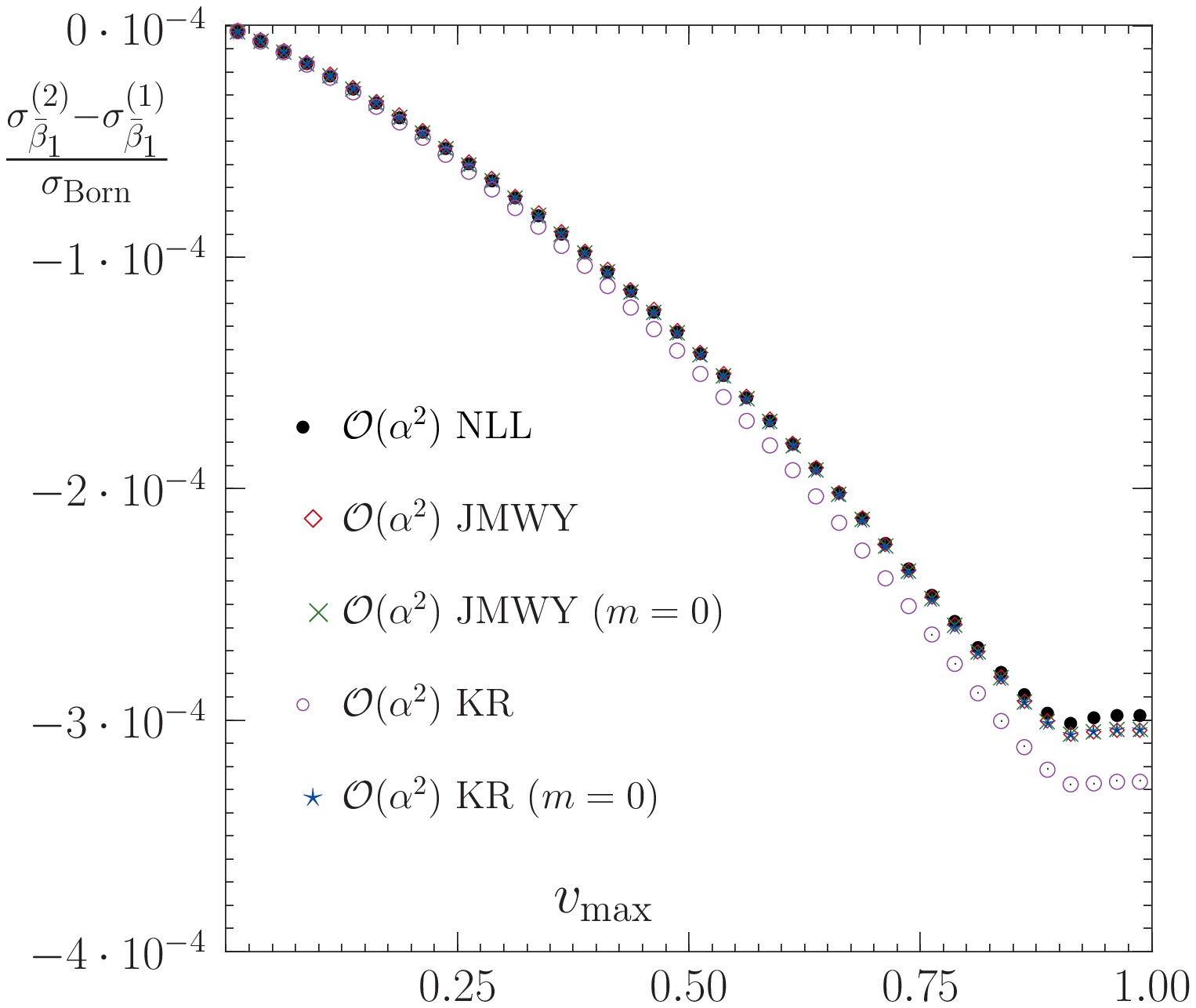}}
\put(3,0){\includegraphics[width=2.75in]{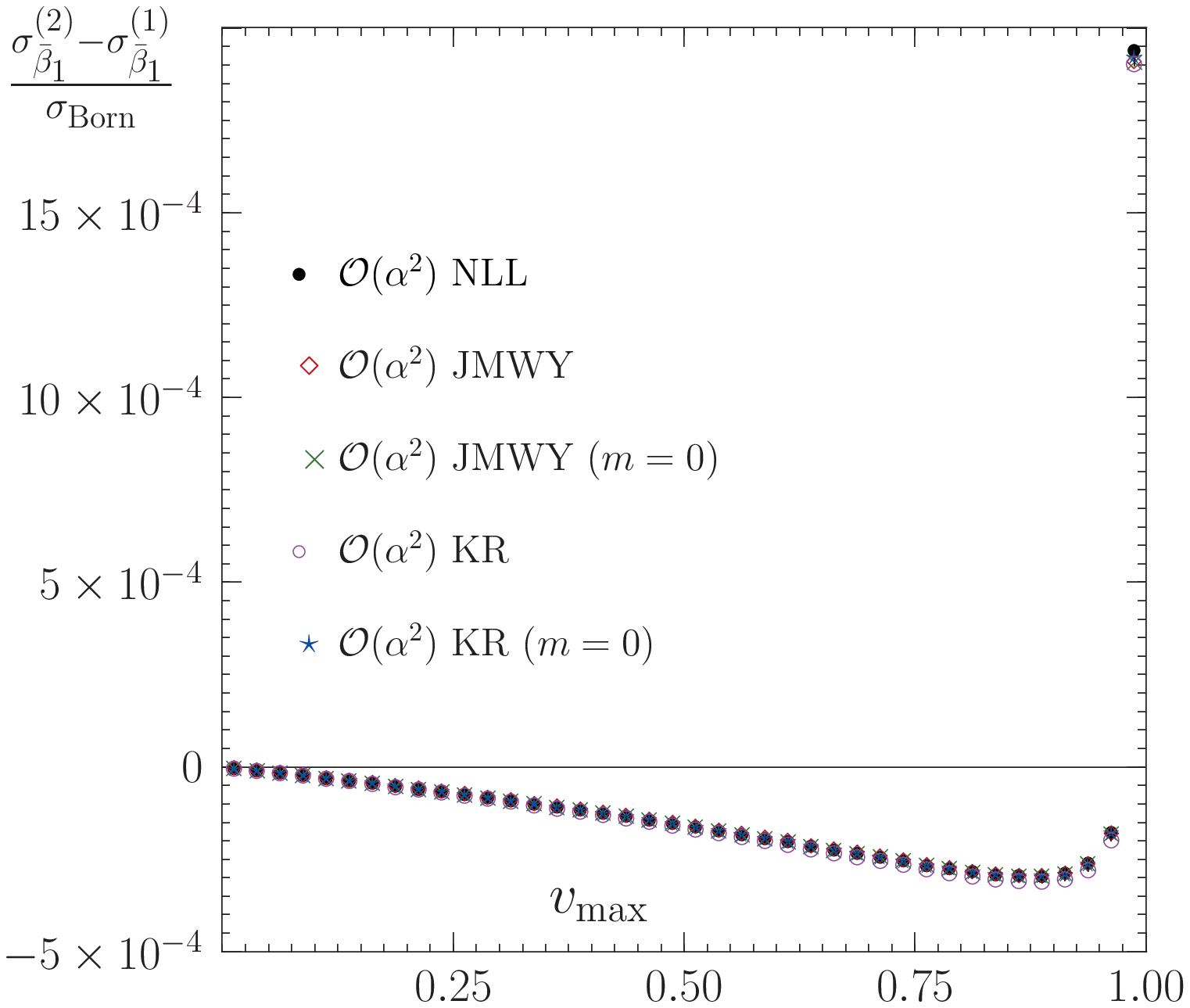}}
\put(1.3,2.6){(a) 1 GeV}
\put(4.2,2.6){(b) 500 GeV}
\end{picture}
\caption{The virtual photon contribution to the $\beta_1$ residual for
muon pair production, integrated up to a cut $v_{\rm max}$. Case (a)
is for a CMS energy of 1 GeV, and case (b) is for a CMS energy of 
500 GeV.}
\end{figure*}

Here, we will restrict our comparisons to the two fully differential
results including mass corrections, JMWY and KR, discussed in the
present paper.  In addition to a 
comparison of the full results, we will compare the ``massless'' limits in
which explicit mass terms in the virtual photon correction are 
omitted in both expressions.  This will be useful in judging the size of
the contribution of the mass corrections alone to the difference in the
cross sections.  Comparisons are made at CMS energies of 1 GeV and 500
GeV.  The 1 GeV scale was chosen as representative of some of the 
radiative return experiments for which PHOKHARA was designed, while the
500 GeV scale anticipates applications at a future linear collider (ILC).
The 1 GeV comparisons are essentially new, since the only previously
reported comparisons at low energy \cite{epiphany} were made before the
matrix elements were completely implemented in their final form: it was
found that some changes were needed to render the results sufficiently
stable numerically when calculated for lower energy scales. The 
expressions shown in the previous sections are adequate for these
calculations.

Figure 1 shows the 
virtual correction $\sigma_{\beta_1}^{(2)} - \sigma_{\beta_1}^{(1)}$, which
is the pure ${\cal O}(\alpha^2)$ contribution to the single hard photon
cross section.  The results are obtained from $10^8$ events created
using the YFS3ff generator, which is the EEX3 option in the \KKMC.
The exact expressions are compared to a NLL approximation
obtained from the expression \Eq{NLL-result}.  All cross sections in
these comparisons are normalized by dividing by the the 
radiationless Born cross-section $\sigma_{\rm Born}$ for 
$e^+ e^- \rightarrow \mu^+ \mu^-$.
It is seen that at both 1 GeV and 500 GeV, the agreement
is very close between all the results compared. 

\begin{figure*}
\label{fig2}
\setlength{\unitlength}{1in}
\begin{picture}(6.5,3.0)
\put(0,0){\includegraphics[height=2.3in]{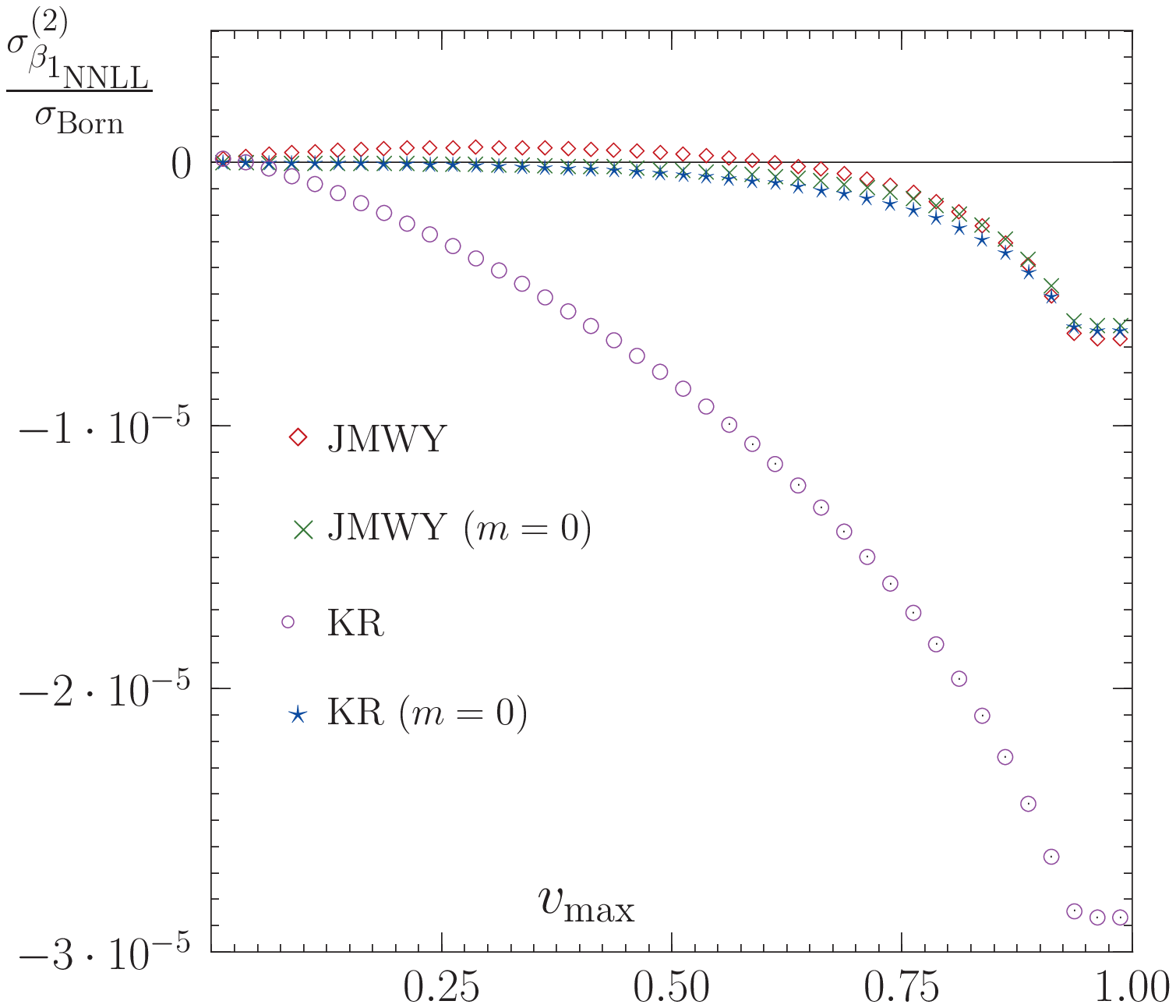}}
\put(3,0){\includegraphics[height=2.3in]{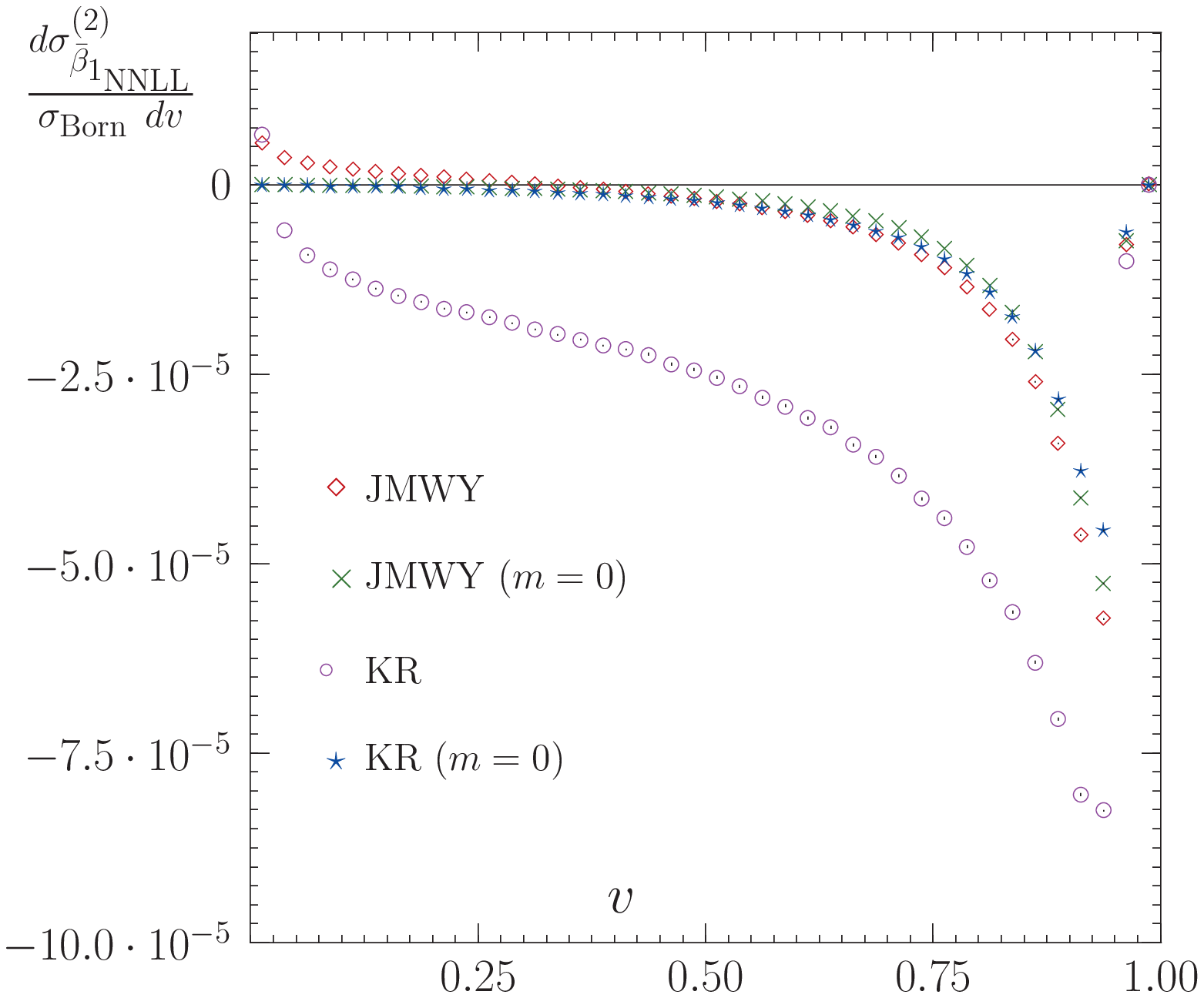}}
\put(1.2,2.6){(a) integrated}
\put(4.1,2.6){(b) differential}
\end{picture}
\caption{
The differences between the exact results and the NLL result are shown
at a CMS energy of 1 GeV both for (a) a cross section integrated to a cutoff
$v_{\rm max}$ and for (b) a partially differential cross section in $v$.
}
\end{figure*}

To see the differences between the various matrix elements, it is 
necessary to look at the NNLL contribution, since we know from the
analytical results in the previous sections that these expressions agree 
to NLL order.  Figure 2 examines the differences between the JMWY and KR 
results and the NLL contribution of \Eq{NLL-result} at 1 GeV.  This is a  
constant contribution (NNLL) to the cross section at ${\cal O}(\alpha^2)$ 
with no factors of the big logarithm $L = \ln(s/m_e^2)$.  In FIG.\ 2(a), the
difference between the integrated YFS residuals is shown, while in
FIG.\ 2(b), the differences $d\sigma_{\bar\beta_1}/dv$ are plotted.
In other words, FIG.\ 2(b) is the derivative of the plot in FIG.\ 2(a),
showing the difference between the virtual corrections at each photon
energy $v$. It is found that the complete results of JMWY and KR
differ by at most $2.2\times 10^{-5}$ units of the Born cross section
when integrated over the full range of $v$.

The differential plot shows
that the difference in the complete virtual correction is essentially 
zero at both endpoints $v = 0, 1$ and reaches a maximum of $3.3\times 10^{-5}$
units of the Born cross section for $v \approx 0.8$. Most of the difference
is found to be due to the mass correction terms. Without these terms, the
results would agree to within at most $10^{-5}$, and much less over most
of the range. The mass corrections agree in both the soft and hard limits
at 1 GeV. The soft agreement mirrors that found analytically in the previous
section. 

\begin{figure*}
\label{fig3}
\setlength{\unitlength}{1in}
\begin{picture}(6.5,3.0)
\put(0,0){\includegraphics[height=2.3in]{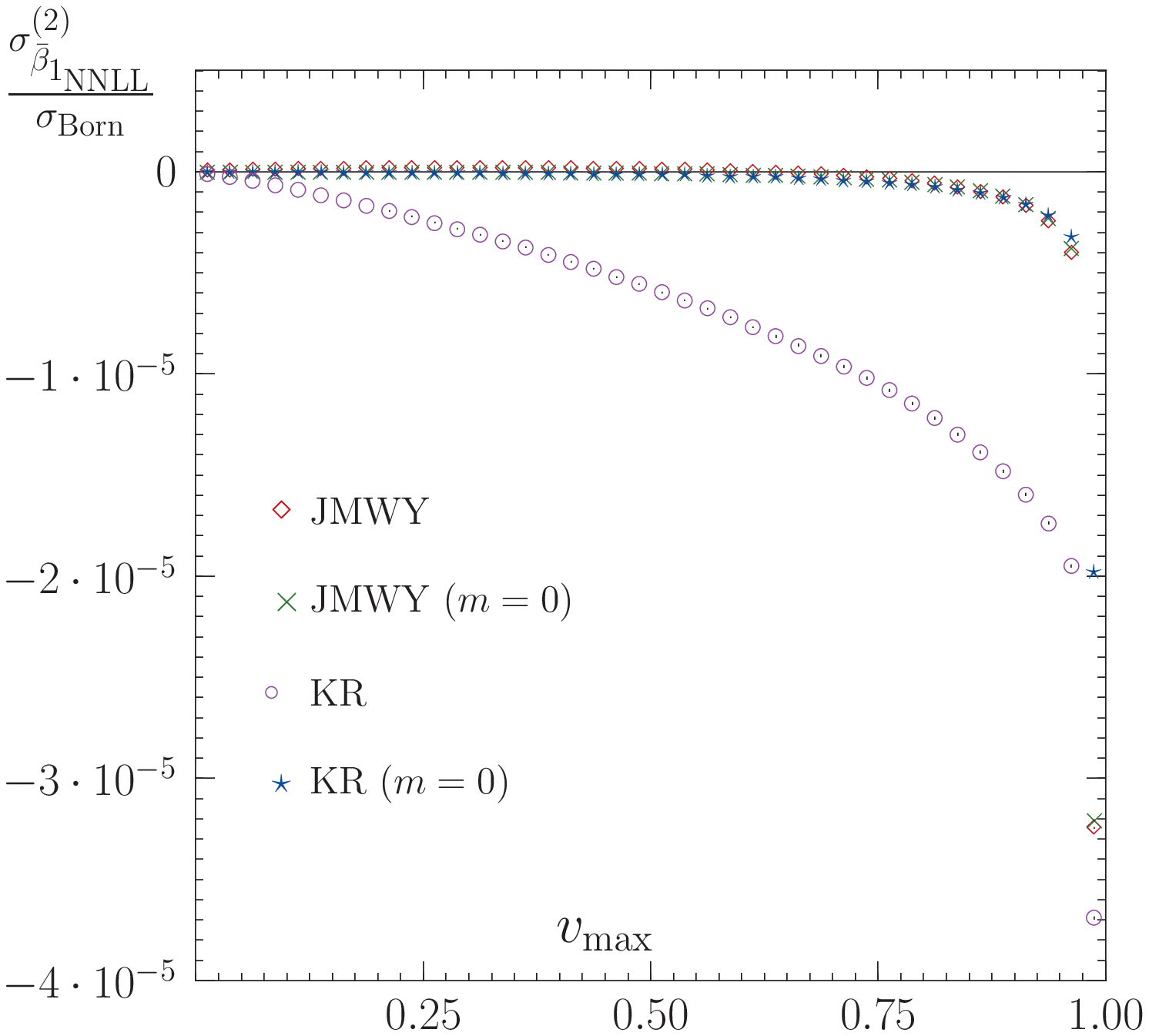}}
\put(3,0){\includegraphics[height=2.3in]{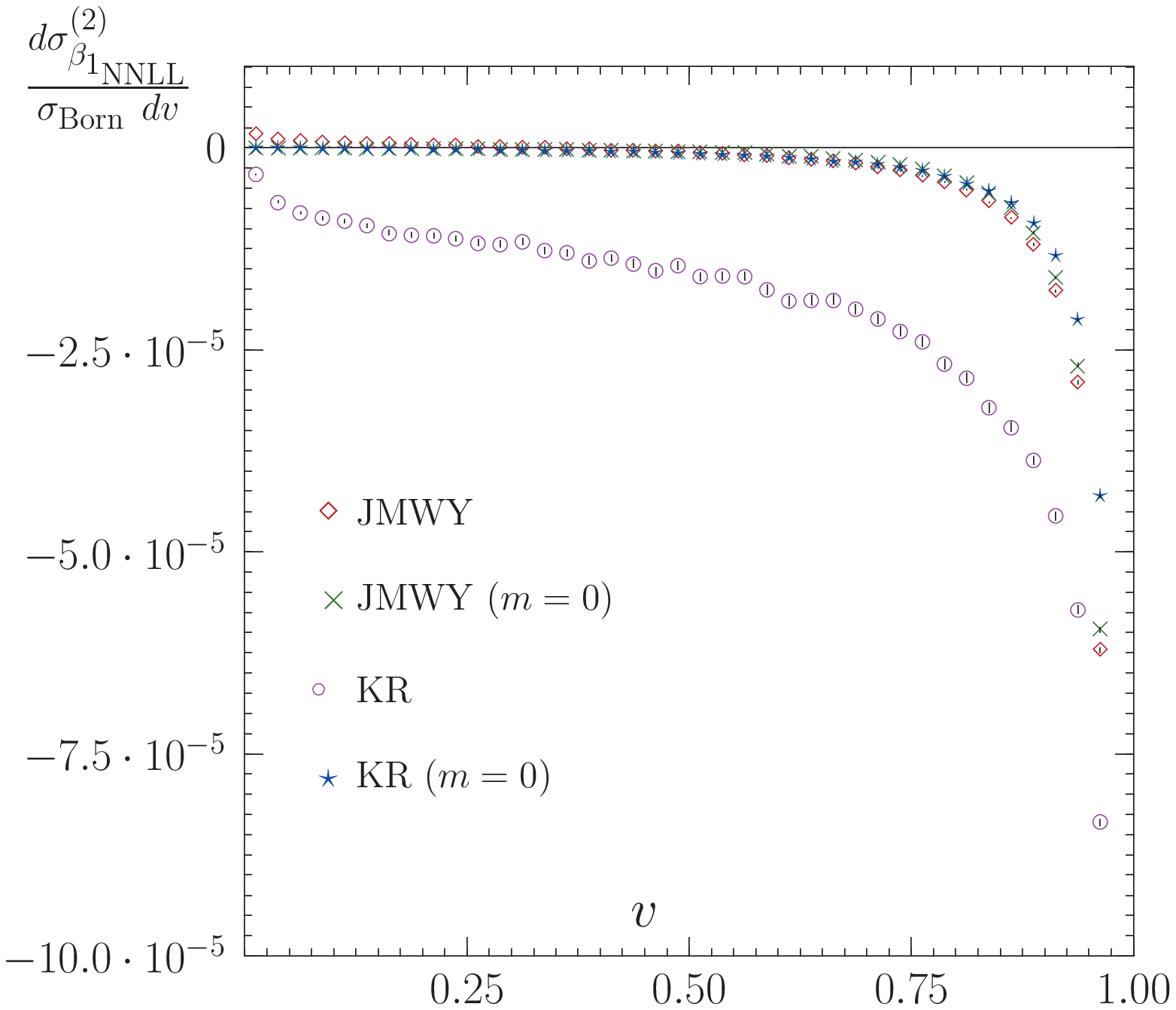}}
\put(1.2,2.6){(a) integrated}
\put(4.1,2.6){(b) differential}
\end{picture}
\caption{
The differences between the exact results and the NLL result are shown
at a CMS energy of 500 GeV both for (a) a cross section integrated to a cutoff
$v_{\rm max}$ and for (b) a partially differential cross section in $v$.
}
\end{figure*}

Figure 3 examines the NNLL contribution to the JMWY and KR results 
at 500 GeV.  FIG.\ 3(a) shows the integrated distributions with the 
NLL contribution from \Eq{NLL-result} subtracted. It is seen that the
complete results, including mass corrections, differ by at most 
$1.6\times 10^{-5}$ units of the Born Cross section.
Again, the bulk of the difference is
seen to be due to the mass corrections. Without these, the results would
agree to within $10^{-6}$, except for the last bin. The last bin shows that
the integrated contribution of the virtual photon factors over the entire
range of photon energies differs by $5\times 10^{-6}$ units of the Born
cross section.

Differential results are compared in FIG.\ 3(b) up to $v = 0.975$. 
The final data point is omitted to permit the rest of the plot to be shown in 
better resolution.  For the last data point ($v = 0.9875$), the KR result was 
$7.0\times 10^{-4}$ for the complete expression and $6.6\times 10^{-4}$ for
the ``massless'' part, while the JMWY result was $11.4\times 10^{-4}$ for
the complete expression and $11.3\times 10^{-4}$ for the massless part.
The difference between the differential results at this point is likely to
be enhanced due the steepness of the distribution. For the remaining points,
the difference is less than $3\times 10^{-5}$. In the limit when $v \rightarrow
1$, additional numerical issues may arise which are separate from the 
collinear singularities, so it is possible that some of the difference in
this limit may be numerical.

The absolute sizes of the NNLL contributions to 
the integals in FIG.\ 2(a) and FIG.\ 3(a) are also of interest, 
since these determine the range
of $v$ for which the much simpler NLL expression can be substituted for the
complete result.  For the JMWY results at 500 GeV, the NNLL contribution
does not exceed $4\times 10^{-6}$ units of the Born cross section for cuts
up to $v_{\rm max} = 0.95$, while for larger $v$, the NLL contribution can
reach the $3.2\times 10^{-5}$ level.  At 1 GeV, the NNLL contribution to
the JMWY result does not exceed $7\times 10^{-6}$ for the full range of 
photon energies. 

The fact that the NNLL contributions to the KR result
are typically larger is mostly due to the mass corrections, which are
considerably greater in the KR expression over most of the range of $v$.
The mass corrections in the JMWY expression reach a level of at most
$10^{-6}$ at 1 GeV, and are smaller at 500 GeV. In contrast, the 
KR mass corrections are typically of the order of $2\times 10^{-5}$ at 
intermediate photon energies. 

While we have expressed the comparisons in units of the non-radiative Born
cross section, it is also useful to express them in units of the integrated
cross section $\sigma^{\rm ISR}_1$ for single real initial-state radiation, 
since there is always at least one photon in the radiative return context.
Integrating the differential cross section of \Ref{BVNB}, 
including the virtual photon contribution which cancels the IR divergence
in the real emission cross section, gives
$\sigma^{\rm ISR}_1 = 0.113\ \sigma_{\rm Born}$ at $\sqrt{s} = 1$ GeV
 and $\sigma^{\rm ISR}_1 = 0.980\ \sigma_{\rm Born}$ at $\sqrt{s} = 500$ GeV.
Dividing by $\sigma^{\rm ISR}_1$ 
instead of $\sigma_{\rm Born}$ would give essentially
the same comparisons at 500 GeV, but increase the differences by a factor
on the order of 10 at 1 GeV. 

\section{Conclusions}
\label{conclusions}

In this paper, we have compared two versions of the exact virtual 
corrections to initial-state bremsstrahlung in fermion pair production in
detail, examining the collinear limits (NLL behavior) analytically both
with and without the explicit mass corrections in each case. We found that
the results of JMWY and KR agree to NLL order, both with and without 
masses, and that both results have the same soft limits. 

Numerical comparisons of the two virtual photon expressions were made by 
integrating them with the {\KKMC}.  The results were presented both
in terms of integrated YFS residuals, to match comparisons made earlier,
and in differential form, to permit comparison of the virtual photon 
corrections as a function of photon energy. It was found that the 
differential results agree to within $3\times 10^{-5}$ units of the
Born cross section across almost all photon energies, for both high energy
(500 GeV) and lower energy (1 GeV) scattering. The difference between the
integrals of the YFS beta functions over the full photon energy range are
also on the order of $10^{-5}$. This level of agreement is compatible with
that found in earlier studies at 
200 GeV.\cite{compare1,paris,beijing,epiphany}.

The two virtual photon expressions compared agree much better when the
mass corrections are omitted from both.  This is because the mass corrections
in the KR expression are typically of order $10^{-5}$, while the JMWY
mass corrections tend to be of order $10^{-6}$ units of the 
Born cross section over most of the photon energy range. Part of this
difference could be due to the structure of the expressions, since the
JMWY expression was created with the expectation of using it in the 
{\KKMC}, while the KR result is foreign to this MC.  However, some 
difference is not surprising, due to the difference in approach to the
mass corrections.  The JMWY result follows the approach of Berends {\it et al}
\cite{berends1}, adding the essential mass corrections for the collinear
limits, and KR use an expansion in powers of $m_e^2/p_i\cdot k$. The results
are very different analytically, with the KR expressions containing many
more terms.

The results found in this paper give a clear estimate of the size of the
difference in two exact, mass-corrected  matrix elements for initial-state 
radiation in fermion pair production.  We have also found that 
these matrix elements agree to NLL order, and have compared two
approaches to adding mass corrections to the matrix element. Such 
information is important
for estimating the precision of these matrix elements in applications to
such processes as the LEP2 final data analyis, radiative return at $\Phi$
and B factories, and anticipated future ILC physics.

\appendix
\section{Stabilized Functions}
\label{appendix}

The stabilized expressions in this paper depend on two sets of 
functions introduced to handle cancellations in differences of logarithms and
dilogarithms with very similar arguments. 
The functions $\Lf_n(x)$ and $\Sf_n(x, y)$ are defined recursively by
\begin{subequations}
\begin{eqnarray}
\label{difdef}
&\Lf_0(x) = \ln(1+x),  \quad &
\Lf_{n+1}(x) = {1\over x}\left(\Lf_n(x) - \Lf_n(0)\right),
\\
&\Sf_0(x, y) = \Sp(x+y),  \quad &
\Sf_{n+1}(x, y) = {1\over y}\left(\Sf_n(x,y) - \Sf_n(x,0)\right).
\end{eqnarray}
\end{subequations}
with $\Sp(x)$ the Spence dilogarithm function, which is also denoted
$\Li_2(x)$. 

Expansions can be used
to evaluate $\Lf_n(x)$ for $x$ small or $\Sf_{n}(x,y)$ for $y$ small.
For the logarithmic difference function $\Lf_n$, 
\begin{equation}
\Lf_n(x) = - \sum_{k=n}^\infty {(-1)^{k}\over k} x^{k-n} .
\end{equation}
Differences between the functions $\Lf_n(x)$ for similar arguments may be 
calculated using the identity
\begin{equation}
{1\over y}\left\{\Lf_n(x+y) - \Lf_n(x)\right\} = {1\over (x+y)^n}\left\{
	{1\over 1+x} \Lf_1\left({y\over 1+x}\right) 
         - \sum_{k=1}^n (x+y)^{k-1}\ \Lf_k(x)\right\}
\end{equation}
valid for $n\ge 1$. Some identities for arguments containing ratios are
\begin{subequations}
\begin{eqnarray} 
(x+y)\ \Lf_1\left({y\over x}\right) &=& x\ \Lf_1\left({-y\over x+y}\right),
\\
(x+y)^2\ \Lf_2\left({y\over x}\right) &=& -x^2\ \Lf_2\left({-y\over x+y}\right)
	- x(x+y),
\\
(x+y)^3\ \Lf_3\left({y\over x}\right) &=& x^3\ \Lf_3\left({-y\over x+y}\right)
	+ {1\over2} xy(x+y).
\end{eqnarray}
\end{subequations}

For the dilogarithmic difference function $\Sf_n$,
\begin{equation}
\Sf_n(x,y) = \sum_{k=n}^\infty {y^{k-n}\over k!} \Sp^{(k)}(x) ,
\end{equation}
where $\Sp^{(k)}(x)$ is the $k^{\rm th}$ derivative of $\Sp(x)$, which
may be calculated recursively using
\begin{eqnarray}
\Sp^{(1)}(x) &=& \Lf_1(-x),
\nonumber\\
\Sp^{(n+1)}(x) &=& {1\over x}\left\{{(n-1)!\over (1-x)^n}
	- n \Sp^{(n)}(x)\right\} \quad\hbox{for}\quad n \ge 1.
\end{eqnarray}
In the limit where both $x$ and $y$ are small, a double expansion is 
useful,
\begin{equation}
\Sf_n(x,y) = \sum_{k=n}^\infty \sum_{l=n}^k {1\over k^2} 
	\left({k\atop l}\right) x^{k-l}  y^{l-n}.
\end{equation}

The functions $\Lf_n(x)$ have a logarithmic singularity at $x = -1$.
The functions $\Sf_n(x,y)$ are singular at
$x = 1$ for $n \ge 1$, but can be calculated in for $x$ approaching 1 with
$x + y < 1$ using the identity
\begin{eqnarray}
\Sf_n(x,y) &=& (-1)^{n+1}\ \Sf_n(1-x,-y)  
    + {1\over x^{n-1}}\ \Lf_1 (-x)\ \Lf_n\left({y\over x}\right) 
\nonumber\\
    &-& {1\over (x-1)^{n-1}}\ \Lf_1(x-1)\ \Lf_n\left({y\over x-1}\right) 
    - {y^q\over x^p (x-1)^p}\ \Lf_p\left({y\over x}\right) \ \Lf_p
	\left({y\over x-1}\right)
\nonumber\\
&+& \sum_{k=1}^{p-1} \left\{ 
    {x^{k-n}\over (1-x)^k}\ \Lf_{n-k}\left({y\over x}\right)
  + (-1)^n {(1-x)^{k-n}\over x^k}\ \Lf_{n-k}\left({y\over x-1}\right)\right\},
\end{eqnarray}
where $n\ge1$, $p$ is the integer part of $(n+1)/2$ and $q = 1$ for $n$ 
even, $q = 0$ for $n$ odd. A useful special case is 
\begin{equation}
\Sf_1(x,y) = \Sf_1(1-x,-y) - \Lf_1(x-1)\ \Lf_1\left({y\over x-1}\right)
    - {1\over x}\ln (1 - x - y)\ \Lf_1\left({y\over x}\right). 
\end{equation}


\begin{thebibliography}{99}
\bibitem{berends}
F.A.\ Berends, W.L.\ Van Neerven, and G.J.H.\ Burgers, { Nucl.\ Phys.}
{\bf B297} (1988) 429 and references therein.

\bibitem{in:1987}
M.\ Igarashi and N.\ Nakazawa, { Nucl. Phys.} {\bf B288}
(1987) 301.

\bibitem{jmwy}
S. Jadach, M. Melles, B.F.L. Ward, and S.A. Yost, { Phys. Rev.} 
{\bf D65} (2002) 073030.

\bibitem{rad6}%with masses
J.H. K\"uhn and G. Rodrigo, { Eur.\ Phys.\ J.}\ {\bf C 25}, 215 (2002).

\bibitem{rad1}
S. Binner, J.H. K\"uhn, and K. Melnikov, { Phys.\ Lett.}\ {\bf B459},
279 (1999).

\bibitem{rad2}
K. Melnikov, F. Nguyen, B. Valeriani, and G. Venanzoni,
{ Phys.\ Lett.}\ {\bf B477}, 114 (2000).

\bibitem{rad3}
H. Czy\.z, J.H. K\"uhn, { Eur.\ Phys.\ J.}\ {\bf C18}, 497 (2001).

\bibitem{rad5}%without masses
G. Rodrigo, A. Gehrmann-De Ridder, M. Guilesaume, and J.H.\ K\"uhn, 
{ Eur.\ Phys.\ J.}\ {\bf C22}, 81 (2001).

\bibitem{rad4}
G. Rodrigo, H. Czy\.z, J.H.\ K\"uhn, and M. Szopa, { Eur.\ Phys.\ J.} 
{\bf C24}, 71 (2002).

\bibitem{hans} 
H. Czyz, A. Grzelinska,
J.H. Kuhn, and  G. Rodrigo, { Eur.\ Phys.\ J.} {\bf C33} (2004) 333.

\bibitem{kkmc:2001}
S.\ Jadach, B.F.L.\ Ward, and Z.\ \Was, { Phys.\ Rev.} {\bf D63} (2001) 
113009;
{ Comput.\ Phys.\ Commun.} {\bf 130} (2000) 260; and references therein. 

\bibitem{hans2}
H. Czyz, A. Grzelinska, J.H.\ Kuhn, and  G. Rodrigo, \textit{SIGHAD03:
Workshop on Hadronic Cross Section at Low Energy, Pisa} (2003),
\hbox{hep-ph/0312217.}

\bibitem{jadcomp}
S. jadach, { Acta Phys.\ Pol.}\ {\bf B36} (2005) 2387.

\bibitem{yfs}
D.R.\ Yennie, S.C.\ Frautschi, and H.\ Suura, { Ann.\ Phys.}
 {\bf 13} (1961) 379;\newline
see also K.T.\ Mahanthappa, { Phys.\ Rev.}\ {\bf 126} (1962) 329, 
for a related analysis.

\bibitem{compare1}
C. Glosser, S. Jadach, B.F.L. Ward, and S. Yost, { Phys.\ Lett.}\
{\bf B605}, 123 (2005).

\bibitem{paris}
S.A. Yost, C. Glosser, S. Jadach, and B.F.L. Ward, \textit{Linear Collider
Workshop 2004}, Paris, \hbox{hep-ph/0409041}.

\bibitem{beijing}
S.A. Yost, C. Glosser, S. Jadach, and B.F.L. Ward, \textit{ICHEP 2004: 
Proceedings of the 32$^{\rm nd}$ International Conference on High Energy 
Physics, Beijing} (World Scientific, Singapore, 2005) 478, 
\hbox{hep-ph/0410238}.

\bibitem{epiphany}
S.A. Yost, S. Jadach, and B.F.L. Ward,  
 { Acta\ Phys.\ Pol.\ }{\bf B36} (2005), 2379.

\bibitem{berends1} 
F.A.\ Berends, R. Kleiss, P. De Causmaecker, R.
Gastmans, W. Troost, and T.T. Wu, { Nucl.\ Phys.}\ {\bf B206}, 61 (1982).

\bibitem{berklei}
F.A.\ Berends, P.\ De Causmaecker, R.\ Gastmans, R.\ Kleiss, W.\ Troost, and
T.T.\ Wu, { Nucl.\ Phys.} {\bf B264} (1986) 243, 265.

\bibitem{xuzhangchang}
Z.\ Xu, D.-H.\ Zhang, and L.\ Chang, { Nucl.\ Phys.} {\bf B291}
(1987), 392.

\bibitem{KS}
R. Kleiss and W.J. Stirling, { Nucl.\ Phys.} {\bf B262} (1985) 235;
Phys. Lett. B{\bf 179} (1986) 159.

\bibitem{form}
J.A.M.\ Vermaseren, \textit{Symbolic Manipulation with FORM} (Computer
Algebra Netherlands, Amsterdam, 1991)

\bibitem{jmwy1}
S. Jadach, M. Melles, B.F.L. Ward, and S.A. Yost, 
	{ Phys.\ Lett.}\ {\bf B377}, 168 (1996).

\bibitem{BVNB} F.A.\ Berends, W.L.\ van Neerven, and G.J.H.\ Burgers, 
{ Nucl.\ Phys.}\ {\bf B297} (1988) 429.
\end{thebibliography}
\end{document}